\documentclass[epj,referee]{svjour}
\usepackage{graphics}

\begin{document}

\title{The reflectivity of relativistic ultra-thin electron layers}
\author{Hui-Chun Wu and J\"{u}rgen Meyer-ter-Vehn}
\institute{Max-Planck-Institut f\"{u}r Quantenoptik, D-85748
Garching, Germany}
\date{Received: date / Revised version:}
\abstract{The coherent reflectivity of a dense, relativistic,
ultra-thin electron layer is derived analytically for an obliquely
incident probe beam. Results are obtained by two-fold Lorentz
transformation. For the analytical treatment, a plane uniform
electron layer is considered. All electrons move with uniform
velocity under an angle to the normal direction of the plane; such
electron motion corresponds to laser acceleration by direct action
of the laser fields, as it is described in a companion paper (paper
I). Electron density is chosen high enough to ensure that many
electrons reside in a volume $\lambda_R^3$, where $\lambda_R$ is the
wavelength of the reflected light in the rest frame of the layer.
Under these conditions, the probe light is back-scattered coherently
and is directed close to the layer normal rather than the direction
of electron velocity. An important consequence is that the Doppler
shift is governed by $\gamma_x=(1-(V_x/c)^2)^{-1/2}$ derived from
the electron velocity component $V_x$ in normal direction rather
than the full $\gamma$-factor of the layer electrons.
  }
\authorrunning{H.-C. Wu and J. Meyer-ter-Vehn}

\maketitle

\begin{figure*}[tbp]
\begin{center}
\resizebox{1.8\columnwidth}{!}{\includegraphics{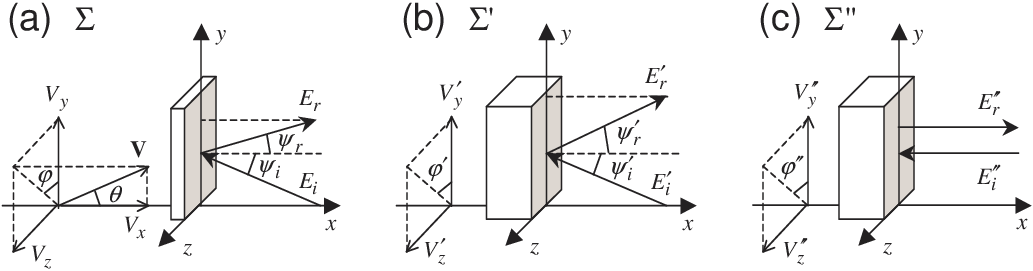}}
\end{center}
\caption{ Different frames and definition of symbols for probe light
obliquely incident in $(x,y)$ plane on a relativistic electron
layer. The uniform layer consists of electrons, having velocity
$\mathbf{V}=[V_x,V_y,V_z]$. (a) laboratory frame $\Sigma$, (b)
intermediate frame $\Sigma^{\prime}$ boosted in $x$-direction such
that $V_x^{\prime}=0$, (c) final fame $\Sigma^{\prime\prime}$,
boosted in $y$-direction such that light is normally incident. }
\label{fig1}
\end{figure*}

\section{Introduction}
\label{1}

High intensity laser pulses have opened new possibilities to create dense electron sheets
moving with velocity $v_m$ close to velocity of light $c$ implying a high relativistic factor
$\gamma=(1-(v_m/c)^2)^{-1/2}\gg 1$. They can act as relativistic mirrors and
Doppler-shift visible light to VUV- and X-ray frequencies by factors $4\gamma_m^2$
with high efficiency \cite{Einstein}. Also the pulse duration is reduced by the
same amount, offering new options for atto-second physics.

Different methods exist to create relativistic mirrors. One way is
to drive plasma waves in a gas jet laser plasma to the threshold of
wave breaking. Non-linear plasma dynamics then form diverging
density spikes moving with the group velocity of the driving laser
pulse \cite{Bulanov2003}. Light reflected from such wake fields has
been detected recently, and $\gamma_x \approx 5$ could be inferred
from the measured Doppler shift \cite{Pirozhkov2007}. Another method is
to reflect intense, linear-polarized laser pulses from solid
surfaces. In this case, the light pressure drives electron
oscillations at the critical surface of the over-dense laser plasma
and forms an oscillating mirror \cite{Lichters}. It generates high
harmonics spectra having universal properties which have been
analyzed in \cite{Gordienko,Baeva}. The surface harmonics have been
observed recently up to photon energies of 3.8 keV \cite{Dromey}. A
third method is to irradiate ultra-thin foils
\cite{Vshivnov,Pirozhkov2006} in a regime in which laser action
separates all electrons from ions forming dense relativistic
electron layers, as it has been described in a companion paper,
referred to as paper I \cite{MtV&Wu}.

The present paper deals with the reflectivity of ultra-thin electron layers
\cite{Kulagin} and is closely related to paper I \cite{MtV&Wu}.
Originating from solid foils, the density of the electron layers
is typically above the critical density.
Nevertheless, they are transparent to optical light
(and even more so to Doppler-shifted light in their rest frame),
because the layer thickness is chosen smaller than the skin depth.
The focus of this paper is on what fraction of probe laser energy is reflected
and how does this reflectivity scale with layer density, thickness,
and $\gamma_m$-factor.

Our particular interest is in coherent reflection. This takes place
when a large number $N$ of electrons resides in a volume $\lambda_R^3$,
where $\lambda_R$ is the wavelength in the rest frame of the layer.
The scattered amplitudes then add coherently, and the reflected signal
scales with $N^2$. This is what also happens in a partially reflecting,
normal mirrors. It is in contrast to incoherent scattering,
which scales $\propto N$ and produces much weaker reflection.
Incoherent scattering from relativistic electrons was discussed in
\cite{Esarey} and has been observed recently also from laser-driven
relativistic electron bunches \cite{Schwoerer}.

As another important difference, coherent emission is oriented along
the normal direction of the relativistic electron layers and
governed by the electron velocity component $V_x$, leading to
Doppler shifts $\propto  \gamma_x^2=1/(1-(V_x/c)^2)$. In contrast,
incoherent emission is oriented along the momentum of individual
scatterers with Doppler shift $\propto \gamma^2$. For layers driven
directly by the laser field, electrons necessarily have large
transverse momentum $p_y$, and
$\gamma_x=\gamma/(1+(p_y/m_ec)^2)^{1/2}$ is typically much smaller
than the full $\gamma$-factor. This leads to significantly reduced
Doppler shifts.

In the following, we derive an analytic expression for coherent
reflection of a probe pulse obliquely incident on a plane, uniform,
relativistic electron layer driven by direct laser action. The
linear reflectivity is considered, i.e. the probe pulse is chosen
weak enough to neglect effects of the probe light on layer dynamics.
Section 2 details two-fold Lorentz transformation into a frame in
which the probe light is normally incident and $V_x=0$. In this
frame, linear coherent scattering is treated in one-dimensional
geometry. The condition for coherent scattering is discussed. In
Section 3, we compare the analytical results with one-dimension
particle-in-cell (1D-PIC) simulations.

\section{Lorentz transformation}
\label{2}

We consider a plane dense electron layer moving at velocity $V_x$ in
normal $x$-direction with relativistic factor
$\gamma_x=(1-(V_x/c)^2)^{-1/2}>1$. Fig. 1a illustrates the geometry
in the laboratory frame $\Sigma$. The uniform layer consists of
electrons, all having the same velocity
\begin{equation}
\mathbf{V}=[V_{x},V_{y},V_{z}]=
V[\cos \theta ,\sin \theta \cos \varphi ,\sin \theta \sin \varphi ].
\end{equation}
Probe light with amplitude $E_i$ is incident from the right-hand side
under an angle $\psi_i$ relative to the normal. The $x,y$ plane is chosen as
plane of incidence; the incident light has wave vector
$\mathbf{k}_{i}=k_{i}[-\cos \psi _{i},\sin \psi _{i},0]$,
and frequency $\omega _{i}=k_{i}c$.

The problem is to determine the amplitude $E_r$, the frequency
$\omega_r$, and the angle $\psi_r$ of the reflected light. In order
to solve the problem, we first perform a Lorentz transformation into
frame $\Sigma^{\prime}$ (see Fig. 1b), moving into $x$-direction
such that $V_x^{\prime}=0$. In a second transformation, we find
frame $\Sigma^{\prime\prime}$ (see Fig. 1c) such that the probe
light is incident normally. This second transformation was proposed
by Bourdier \cite{Bourdier} to treat oblique incidence in
one-dimensional laser plasma configurations. We now give the
transformation formulas explicitly, before treating the 1D
reflection problem in Section 3.

\subsection{$\Sigma $ $\rightarrow $ $\Sigma ^{\prime }$}
\label{2.1}

Relative to the lab frame $\Sigma $, the intermediate frame
$\Sigma ^{\prime }$ moves with a longitudinal velocity $V_{x}\mathbf{e}_{x}$,
where $\mathbf{e}_{x}$ is the unit vector in the $x$ direction.
In $\Sigma ^{\prime }$, the mirror has only transverse velocity $\mathbf{V}%
^{\prime }=\gamma _{x}[0,V_{y},V_{z}]$, where
$\gamma _{x}=(1-\beta_{x}^{2})^{-1/2}$ and $\beta _{x}=V_{x}/c$.
Its relativistic factor becomes $\gamma ^{\prime }=\gamma /\gamma _{x}$.
Thickness and density of the mirror are
$d^{\prime }=d\gamma _{x}$ and $n_{e}^{\prime }=n_{e}/\gamma _{x}$,
obtained from the transformation of the 4-vector
[$n_{e}\mathbf{V},n_{e}c$].

In $\Sigma ^{\prime }$, the field amplitude is
$E_{i}^{\prime }=E_{i}\gamma _{x}(1+\beta _{x}\cos \psi _{i})$,
independent of the polarization direction.
The incident light has frequency
$\omega _{i}^{\prime }=k_{i}^{\prime }c=\omega _{i}\gamma _{x}(1+\beta_{x}\cos \psi _{i})$
and wave vector
$\mathbf{k}_{i}^{\prime}=k_{i}[-\gamma _{x}(\cos \psi _{i}+\beta _{x}),\sin \psi _{i},0]$.
>From this one obtains the angle of incidence
\begin{equation}
\tan \psi _{i}^{\prime }=\frac{\sin \psi _{i}}{\gamma _{x}(\cos \psi
_{i}+\beta _{x})}.
\end{equation}

\subsection{$\Sigma ^{\prime }$ $\rightarrow $ $\Sigma ^{\prime\prime}$}
\label{2.2}
Relative to $\Sigma ^{\prime }$, the frame
$\Sigma^{\prime\prime }$ moves with a transverse velocity
$c\sin \psi _{i}^{\prime }\mathbf{e}_{y}$.
This special transformation is widely used to implement oblique
incidence in 1D PIC codes \cite{Bourdier}. It transforms obliquely incident
into normally incident light.
In the frame $\Sigma ^{\prime \prime }$, the electron velocity is
\begin{equation}
\mathbf{V}^{\prime \prime }=[0,V_{y}^{\prime }-c\sin \psi _{i}^{\prime },
V_{z}^{\prime }\cos \psi _{i}^{\prime }]
/(1-\beta _{y}^{\prime }\sin\psi _{i}^{\prime }),
\end{equation}
and one has
$\gamma ^{\prime \prime }=\gamma ^{\prime}
(1-\beta _{y}^{\prime }\sin \psi _{i}^{\prime })
/\cos \psi_{i}^{\prime }$.
The electron density becomes
$n_{e}^{\prime \prime}=n_{e}^{\prime }(1-\beta _{y}^{\prime }
\sin \psi_{i}^{\prime })/\cos \psi _{i}^{\prime }$.
The layer thickness is $d^{\prime \prime }=d^{\prime }$.

For the incident light, we have
$E_{i}^{\prime \prime}=E_{i}^{\prime }\cos \psi _{i}^{\prime }$,
$\omega _{i}^{\prime \prime }=\omega _{i}^{\prime }\cos \psi_{i}^{\prime }$,
and
$\mathbf{k}_{i}^{\prime \prime}=[-k_{i}^{\prime }\cos \psi _{i}^{\prime },0,0]$.
So the angle of incidence is $\psi _{i}^{\prime \prime }=0$.
It is noted that there are two Lorentz invariants
$E_{i}^{\prime\prime }/\omega _{i}^{\prime \prime }$ $=E_{i}^{\prime }
/\omega_{i}^{\prime }=E_{i}/\omega _{i}$
and
$n_{e}^{\prime \prime }/\gamma^{\prime \prime }=n_{e}^{\prime }
/\gamma ^{\prime }=n_{e}/\gamma $,
which is the electron density in the rest frame of the mirror.

As we shall derive below, the reflected signal has
$\mathbf{k}_{r}^{\prime \prime }=[k_{i}^{\prime }\cos \psi
_{i}^{\prime },0,0]$, $\omega _{r}^{\prime \prime }=\omega
_{i}^{\prime \prime }$, and $\psi _{r}^{\prime \prime }=0$. Its
amplitude $E_{r}^{\prime \prime }$ will be calculated in Sec. 3. The
polarizations of $E_{r}^{\prime \prime }$ and $E_{i}^{\prime \prime
}$ are the same.

\subsection{$E_{r}^{\prime\prime}\rightarrow E_{r}^{\prime}$}
\label{2.3}

In frame $\Sigma ^{\prime }$, we have $\omega _{r}^{\prime }=\omega _{i}^{\prime }$,
$\mathbf{k}_{r}^{\prime }=k_{i}^{\prime }[\cos \psi _{i}^{\prime },
\sin \psi_{i}^{\prime },0]$, and $\psi_{r}^{\prime }=\psi _{i}^{\prime }$.
This means the reflected wave has the same frequency as the incident light
and the reflection angle is equal to the angle of incidence. The field amplitude is
$E_{r}^{\prime}=E_{r}^{\prime\prime}/\cos \psi _{i}^{\prime }$.

\subsection{$E_{r}^{\prime }\rightarrow E_{r}$}
\label{2.4}
In the lab frame, the reflected signal has frequency
$\omega _{r}=\omega_{i}^{\prime }\gamma _{x}(1+\beta _{x}\cos \psi
_{i}^{\prime })$ and wave vector $\mathbf{k}_{r}=k_{i}^{\prime}
[\gamma _{x}(\cos \psi _{i}^{\prime }+\beta _{x}),\sin
\psi_{i}^{\prime },0]$. Using $\omega _{i}^{\prime }$ obtained in
Sec. 2.1, one gets the Doppler shift

\begin{equation}
\omega _{r}=\omega _{i}\gamma _{x}^{2}(1+\beta _{x}\cos \psi _{i})(1+\beta
_{x}\cos \psi _{i}^{\prime }).
\end{equation}
>From the wave vector $\mathbf{k}_{r}$ one finds the reflection angle $\psi _{r}$
in the form
\begin{equation}
\tan \psi _{r}=\frac{\sin \psi _{i}^{\prime }}{\gamma _{x}
(\cos \psi_{i}^{\prime }+\beta _{x})},
\end{equation}
where the incidence angle $\psi _{i}^{\prime }$ in $\Sigma ^{\prime
}$ is given by Eq. (2). Finally the field amplitude satisfies
$E_{r}=E_{r}^{\prime}\gamma _{x}(1+\beta _{x}\cos \psi _{i}^{\prime
}) =E_{r}^{\prime \prime}\gamma _{x}(\sec \psi _{i}^{\prime }+\beta
_{x})$. Again $E_{r}^{\prime\prime}/\omega _{r}^{\prime\prime}
=E_{r}^{\prime }/\omega_{r}^{\prime }=E_{r}/\omega _{r}$ is a
Lorentz invariant.

As an important result, we find that, seen from the lab frame
$\Sigma$, the light is reflected into an angle $\psi_r$ much smaller
than the angle of incidence $\psi_i$, depending on $\gamma _{x}$.
This is shown in Fig. 2a, where $\psi _{r}$ is plotted as a function
of $\gamma _{x}$ for $\psi_i=30{\textordmasculine}$ and
$60{\textordmasculine}$. With the increase of $\gamma _{x}$, the
reflected light turns strongly into the normal direction of the
layer. For comparison, a layer at rest ($\gamma_x=1$) would give
simple specular reflection with $\psi _{i}=\psi _{r}$. Also the
Doppler shift given by Eq. (4) depends on $\beta_x$ and
$\gamma_x=\sqrt{1-\beta_x^2}$ rather than the full $\mathbf{\beta}$
and $\gamma$ of the individual electrons. Only for the special case
$\gamma_x=\gamma$, the full Doppler shift
$\omega_r/\omega_i=\gamma^2(1+\beta)^2$ can be reached for normal
incidence. For oblique incidence, $\omega_r/\omega_i$ is plotted as
function of $\psi_i$ in Fig. 2b for $\gamma_x=\gamma=5$.

The present result holds for coherent scattering from a dense
electron layer and is in distinct contrast to incoherent Thomson
scattering which scatters the light into a direction close to that
of the individual electron momenta. Even though all the relativistic
electrons in the case studied here move into the same direction,
their angle $\theta$ relative to the normal $x$-direction (see Fig.
1a) plays no apparent role in the coherent scattering process. The
reason for this will become clearer in section 3, where we study the
scattering itself in frame $\Sigma^{\prime \prime}$.

Lorentz transformations of p- and s-polarized light fields always
lead to p- and s-polarized reflected fields, respectively, so the
polarization states of incident and reflected wave remain the same.
In the normal-incidence frame $\Sigma ^{\prime\prime }$, there is no
difference between p- and s-polarized fields, so that the field
ratios $E_{r}^{\prime \prime }/E_{i}^{\prime \prime }$ are same for
both cases. Since field amplitude transformations are same for both
polarization states, also the field ratio $E_{r}/E_{i}$ in the lab
frame is independent of the polarization states.

\begin{figure*}[tbp]
\begin{center}
\resizebox{1.5\columnwidth}{!}{\includegraphics{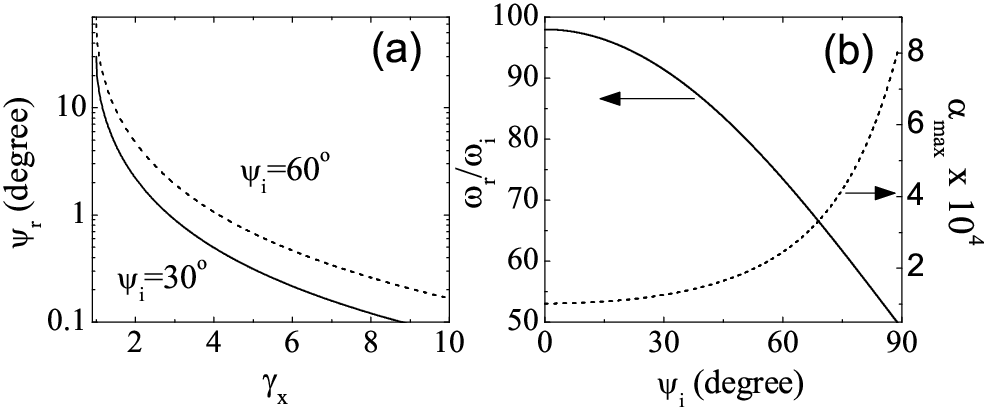}}
\end{center}
\caption{ (a) The reflection angle $\protect\psi_{r}$ as a function
of $\protect\gamma _{x}$ for two angles of incidence, $\psi_i=
30\textordmasculine$ and $60\textordmasculine$. (b) Frequency
$\protect\omega _{r}$ and maximum reflectivity $\protect\alpha
_{\max }$ as a function of the angle of incidence $\protect\psi
_{i}$. In both cases, it is set $\protect\gamma _{x}=\protect\gamma
=5$. }\label{fig2}
\end{figure*}

\section{Coherent Thomson scattering}
\label{3}

We are now in a position to calculate in frame $\Sigma^{\prime \prime}$
what fraction of incident light is reflected from the relativistic electron layer
by coherent Thomson scattering. The layer has thickness $d^{\prime \prime}$
and electron density $n_e^{\prime \prime}$. The condition for coherent scattering
is that many electrons reside in a volume $\lambda_{i}^{\prime \prime 3}$,
i.e. $n_{e}^{\prime \prime }\lambda _{i}^{\prime\prime 3}\gg 1$.
For the simple case of $\psi _{i}=0$, this is equivalent to the condition
\begin{equation}\label{coherence condition}
n_{e}\gg 10^{13}\gamma_{x}^{4}cm^{-3}
\end{equation}
in lab frame $\Sigma$. For $\gamma _{x}=10$ and $100$, coherent scattering
requires $n_{e}\gg 10^{17}$cm$^{-3}$ and $10^{21}$cm$^{-3}$, respectively.


Let us now determine the reflectivity of such layers. For this we have
to calculate the reflected amplitude $E_{r}^{\prime\prime}$ from the incident
amplitude $E_{i}^{\prime\prime}$ for normal incidence in frame
$\Sigma^{\prime\prime}$. The result has then to be transformed back
to the lab frame $\Sigma$ to obtain the reflection coefficient
of photon energy and number. We start from the wave equation
for the reflected light in plane 1D geometry
\begin{equation}\label{wave quation}
(\frac{\partial ^{2}}{\partial x^{2}}-\frac{1}{c^{2}}\frac{\partial ^{2}}{%
\partial t^{2}})A_{r}^{\prime \prime}(x,t)
=S^{\prime \prime}(x,t)=-\mu _{0}J^{\prime \prime}(x,t).
\end{equation}
Here $A_{r}^{\prime \prime }$ is the vector potential, and
$S^{\prime \prime}$ is the source of the scattered radiation. It is
given by the current $J^{\prime \prime}(x,t)$. We note that, in Eq.
(7), the coordinates $(x,t)$ refer also to frame
$\Sigma^{\prime\prime}$, but for simpler notation we have dropped
the double prime index. The incident laser field $E_{i}^{\prime
\prime }(x,t)$ $=E_{i0}^{\prime \prime }\sin (k_{i}^{\prime \prime
}x+\omega_{i}^{\prime \prime }t)$ induces the linear current
\begin{equation}
J^{\prime\prime}=-[e^{2}E_{i0}^{\prime \prime }n_{e}^{\prime \prime }(x)
/\gamma^{\prime \prime }m_{e}\omega _{i}^{\prime \prime }]
\cos(k_{i}^{\prime \prime }x+\omega _{i}^{\prime \prime }t),
\end{equation}
where $n_{e}^{\prime \prime }(x)$ is the electron density,
$\gamma ^{\prime \prime }$ the relativistic factor (see Sec. 2.2),
$m_{e}$ the rest mass, and $-e$ the charge of the electron.
Notice that the uniform direct current related to the transverse
velocities $V_{y}^{\prime \prime }$ and $V_{z}^{\prime \prime }$ is time-independent
in the present model and does not contribute to the scattered radiation.
The full electron velocity and its $\gamma$ factor including the transverse component
enters the source term of Eq. (\ref{wave quation}) only through
the relativistic mass $\gamma^{\prime \prime }m_e$.
Also we consider here weak probe pulses only with amplitude
$a_{0}=eE_{i0}^{\prime\prime }/m_{e}c\omega _{i}^{\prime \prime }<1$,
which is the same in all frames.
The small $a_{0}$ has negligible influence on $\gamma^{\prime \prime }$.
Effects of non-linear Thomson scattering are not considered.

Another important assumption is made in deriving the current and
should be emphasized. In taking a smooth electron density function
$n_{e}^{\prime \prime }(x)$, rather than a random distribution of
point electrons, we exclude incoherent scattering and suppose
implicitly that the condition (\ref{coherence condition}) for
coherent scattering is fulfilled. In order to study the transition
from incoherent to coherent scattering, the distribution of the
individual scatterers has to be taken into account explicitly. This
transition may become relevant in experiments, since the density
profile of the layers decays in time. We expect that this will lead
to significant changes in the scattered spectra with emission
direction turning from normal $x$-direction to the direction of
electron momenta and Doppler shifts changing from $4\gamma_x^2$ to
$4\gamma^2$ scaling.

Equation (7) has the solution
\begin{equation}
A_{r}^{\prime \prime }(x,t)={\int\!\!\!\!\int G(x-x^{\prime },t-}t^{\prime }%
{)S(x^{\prime },t^{\prime })dx^{\prime }d}t^{\prime },
\end{equation}
where $G(x-x^{\prime },t-t^{\prime })
=-\frac{c}{2}H[(t-t^{\prime})-|x-x^{\prime }|/c]$ is the Green
function of Eq. (7), and $H$ is the Heaviside step function.
Equation (9) leads to
\begin{equation}
A_{r}^{\prime \prime }(x,t)=-\frac{c}{2}\int_{-\infty }^{\infty }dx^{\prime
}\int_{-\infty }^{t-|x-x^{\prime }|/c}S(x^{\prime },t^{\prime })dt^{\prime }.
\end{equation}
>From Eq. (10), one obtains the electric field
\begin{eqnarray}
E_{r}^{\prime \prime } &=&-\frac{\partial A_{r}^{\prime \prime }}{\partial t}
=\frac{c\mu _{0}e^{2}E_{i0}^{\prime \prime }}{2\gamma ^{\prime\prime }
m_{e}\omega _{i}^{\prime \prime }}\nonumber \\
&&\times  \int_{-\infty }^{\infty}n_{e}^{\prime \prime }\cos
[k_{i}^{\prime \prime } x^{\prime}+\omega _{i}^{\prime \prime
}(t-|x-x^{\prime }|/c)]dx^{\prime}.
\end{eqnarray}
For simplicity, we consider an electron slab with uniform density
$n_{e}^{\prime\prime }(x)=n_{e0}^{\prime \prime }H(-x)H(x+d^{\prime
\prime })$. For the reflected wave propagating in $+x$ direction,
Eq. (8) leads to

\begin{equation}
E_{r}^{\prime \prime }(x,t)=\frac{e^{2}n_{e0}^{\prime \prime }E_{i0}^{\prime
\prime }}{2m_{e}\varepsilon _{0}\gamma ^{\prime \prime }\omega _{i}^{\prime
\prime 2}}\sin \xi \cos (\omega _{i}^{\prime \prime }t-k_{i}^{\prime \prime
}x-\xi ),
\end{equation}
where $\xi \equiv k_{i}^{\prime \prime }d^{\prime \prime }$.
Making use of the relations derived in Section 2, the field
$E_{r}^{\prime \prime }(x,t)$ is now transformed back into the
lab frame, where
$E_{r}(x,t)=E_{r0}\cos (\omega _{i}t-k_{i}x-k_{i}^{\prime \prime }d^{\prime\prime })$
is obtained with the amplitude
\begin{equation}
E_{r0}=E_{i0}\frac{\omega _{p0}^{2}}{2\omega _{i}^{2}}\frac{(\sec \psi
_{i}^{\prime }+\beta _{x})\sin \xi }{\gamma (1+\beta _{x}\cos \psi _{i})\cos
\psi _{i}^{\prime }}.
\end{equation}%
Here $\omega _{p0}=\sqrt{e^{2}n_{e0}/m_{e}\epsilon _{0}}$ is the plasma frequency,
and $\xi =k_{i}d\gamma _{x}^{2}(1+\beta _{x}\cos \psi _{i})\cos\psi _{i}^{\prime }$.

Taking into account the ratio of pulse durations of incident and
reflected wave which is given by $\tau _{r}/\tau _{i}=\omega
_{i}/\omega _{r}$, Eq. (13) allows to calculate the fraction of
pulse energy reflected by the relativistic electron layer:

\begin{equation}
\alpha =\frac{E_{r0}^{2}\tau _{r}}{E_{i0}^{2}\tau _{i}}=\frac{\omega_{p0}^{4}}
{4\omega _{i}^{4}}\frac{(1+\beta _{x}\cos \psi _{i}^{\prime })\sin^{2}\xi }
{\gamma ^{2}\gamma _{x}^{2}(1+\beta _{x}\cos \psi _{i})^{3}\cos^{4}\psi _{i}^{\prime }}.
\end{equation}%
This is the central result of the present paper. We call $\alpha $
the reflectivity for coherent Thomson scattering. Notice that
$\alpha \propto \omega_{p0}^{4}/\gamma^2 \propto n_{e0}^{2}/(\gamma
m_e)^2$. Apparently, it depends quadratically on the layer density,
as it should be for coherent scattering, and on the full $\gamma$
factor exclusively through the relativistic mass $\gamma m_e$.
Another remarkable feature is that the direction of electron motion
defined by the angle $\theta$ plays no special role. In contrast to
incoherent scattering, the coherently scattered radiation exhibits
no conspicuous maxima when choosing $\psi_i=\theta$ and $\phi=\pi$
(see discussion about $\varphi$ below), in which case the probe
radiation is incident just head-on to the electron momenta.

\begin{figure*}[htp]
\begin{center}
\resizebox{1.5\columnwidth}{!}{\includegraphics{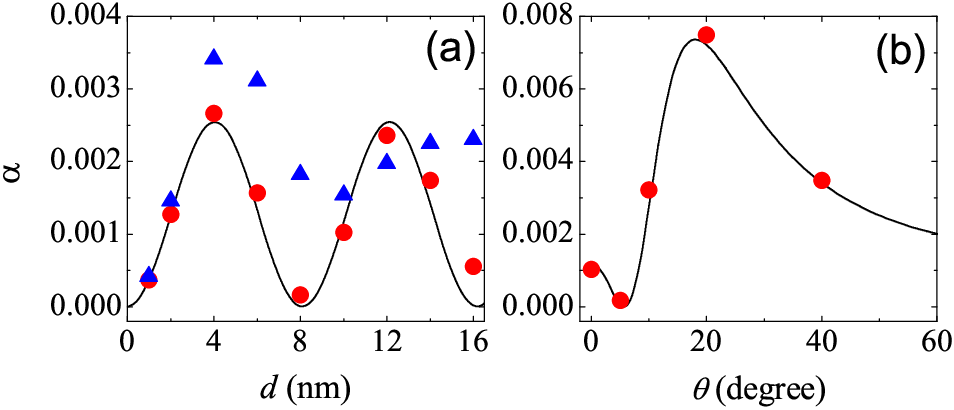}}
\end{center}
\caption{(Color online) Reflectivity $\protect\alpha$ for
$\protect\gamma=5$, $\protect\psi_i=0$, and $n_{e0}=5n_{c}$. The
analytical result (solid line, Eq. (15)) is compared with PIC
simulations: (a) dependence on layer thickness $d$ for
$\protect\gamma_x=\gamma=5$ ($\protect\theta=0$), (b) dependence on
electron direction $\protect\theta$ for fixed $d=10$ nm. Few-cycle
cosine-type pulses were used in the simulation with duration
$T=8\protect\tau _{i}$ (circles) and $T=3\protect\tau _{i}$
(triangles). } \label{fig3}
\end{figure*}

Instead $\alpha$ depends explicitly on the normal velocity component
$\beta_x$, $\gamma_x=(1-\beta_x^2)^{-1/2}$, and kinematic factors
related to oblique incidence. This is also true for the factor
$\sin^2\xi=\sin^2(k_i^{\prime\prime}d^{\prime\prime})=\sin^2(\pi
d/L)$, describing oscillations of the reflectivity as a function
layer thickness $d$. The period of these oscillations is given by
$L=\lambda _{i}/[2\gamma _{x}^{2}(1+\beta_{x}\cos \psi _{i}) \cos
\psi _{i}^{\prime }]\approx \lambda _{r}$, i.e. the wavelength of
the reflected radiation in the lab frame. The periodic modulation of
$\alpha $ with $d$ is due to coherent superposition of radiation
scattered from different depths of the electron layer. It is shown
for normal incidence and $\gamma _{x}=\gamma =5$ in Fig. 3a. The
first maximum shows up for $d=\lambda _{r}/2$ and minima at
$d=\lambda _{r},2\lambda _{r},...$ . In Fig. 2b, the maximum
reflectivity $\alpha _{\max }$ is plotted versus angle of incidence
$\psi _{i}$; for the oblique incidence, the reflectivity increases,
while the frequency of the reflected radiation decreases.

We also point out that the reflection frequency $\omega _{r}$, angle
$\psi _{r}$ and reflectivity $\alpha $ are all independent of the
azimuthal angle $\varphi $, i.e. the direction of the transverse
velocity of the mirror. In Fig. 1a, one can simply set $\varphi$ to
zero or $\pi$. However, the azimuthal angle $\varphi $ is contained
in $n_{e}^{\prime \prime }$, which has an influence on the criterion
in Eq. (6), provided that $\psi _{i}\neq 0$.

For normal incidence with $\psi _{i}=0$, expression (14) for
$\alpha$ reduces to
\begin{equation}
\alpha (\psi _{i}=0)=\frac{\omega _{p0}^{4}}{4\omega _{i}^{4}}\frac{\sin
^{2}[\gamma _{x}^{2}(1+\beta _{x})k_{i}d]}{\gamma ^{2}\gamma
_{x}^{2}(1+\beta _{x})^{2}}.
\end{equation}

The reflection coefficient of photon number is a Lorentz invariant and is
easily obtained in the form
\begin{equation}\label{alpha_ph}
\alpha _{ph}=\alpha \frac{\omega _{i}}{\omega _{r}}=\frac{\omega _{p0}^{4}}{%
4\omega _{i}^{4}}\frac{\sin ^{2}\xi }{\gamma ^{2}\gamma _{x}^{4}(1+\beta
_{x}\cos \psi _{i})^{4}\cos ^{4}\psi _{i}^{\prime }}.
\end{equation}%
In the limit $\psi _{i}=0$ and $\xi \ll 1$, we find
\begin{equation}
\alpha_{ph}\approx (n_{e0}k_{i}d/4n_{c}\gamma )^{2},
\end{equation}
where $n_{c}$ is the
critical density corresponding to frequency $\omega_i$. This result
is also contained in Ref. \cite{Bulanov2003}. One should realize,
however, that the condition $\xi \ll 1$, i.e. $d\ll \lambda_{r}/\pi
$, is difficult to fulfil experimentally, in particular for higher
values of $\gamma_{x}$.

\begin{figure*}[tbp]
\begin{center}
\resizebox{2\columnwidth}{!}{\includegraphics{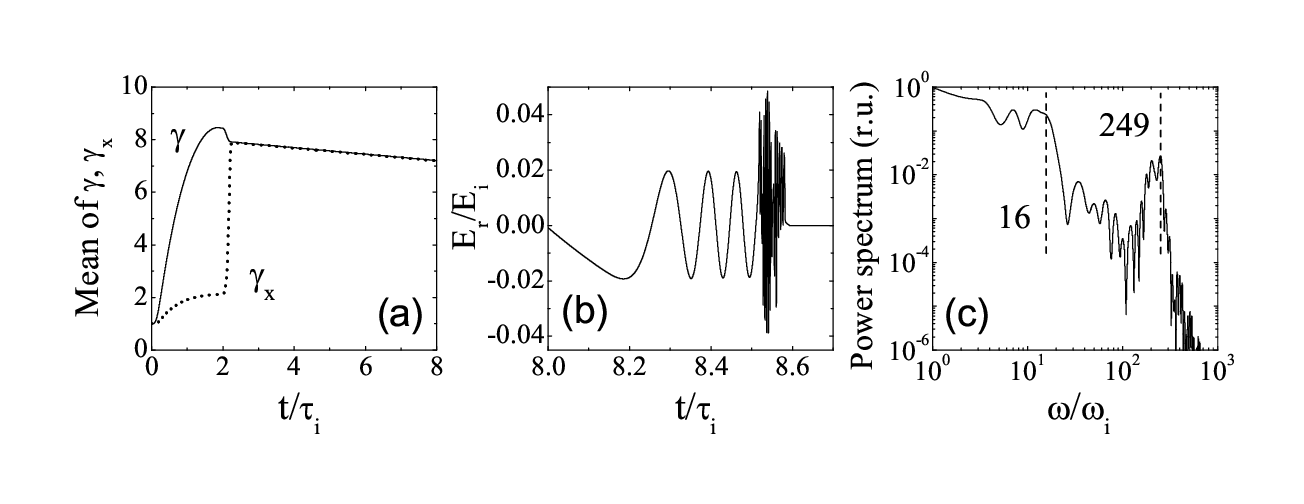}}
\end{center}
\caption{Results of PIC simulation of probe laser ($a_p=0.01$) reflection
from ultrathin foil ($d=5$ nm, $n_e/n_c=1$), irradiated by drive laser pulse
with $a_d=2$ and half-cycle switch pulse.
(a) Time evolution of $\protect\gamma $ and $\protect\gamma _{x}$ averaged
over electron layer. (b) Reflected signal. (c) Corresponding power spectrum.
For details see text.}
\label{fig4}
\end{figure*}


\section{PIC simulation}
\label{PIC}

We have performed 1D PIC simulation to check the mirror reflectivity $\alpha$ derived above.
For a direct comparison, we have introduced an ion background in the simulation,
moving synchronously with the
electron layer to suppress Coulomb expansion. This charge-compensating background has almost
no effect on the Thomson scattering because of the high ion mass.
In the simulation, the probe laser pulse is perpendicularly incident ($\psi=0$) and
has amplitude $a_{0}=eE_{i0}/m_{e}c\omega _{i}=0.01$, wavelength $\lambda _{i}=800$ nm,
and a $\sin ^{2}(t/T)$ envelope. Two different pulse durations,
$T=8\tau _{i}$ and $T=3\tau _{i}$, were used.

Comparisons with the analytical results are shown in Fig. 3. In Fig.
3a, $\alpha$ is plotted versus layer thickness $d$ for fixed
$\gamma_x=\gamma=5$ ($\theta=0$). The modulation due to constructive
and destructive interference is well reproduced by the simulated
results for the longer pulse duration (circles), but for the short
($T=3\tau _{i}$, triangles) pulse agreement is found for $d \ll 4$ nm
below the first maximum of $\alpha$. The reason is incomplete
destructive interference for very short pulses due to time delay of
reflected signals from deeper layers. It is found that the
reflectivity saturates for thicker layers at almost the level of the
first maximum. In Fig. 3b, we kept $d=10$ nm and $\gamma =5$ fixed
and varied the inclination angle $\theta$ of electron momenta
relative to $x$ direction. Again, we find perfect agreement between
simulated long pulse results and Eq. (15).

Finally, we compare with a more realistic PIC simulation
(no artificial ion background). In this case, we have to extract
space- and time-averaged density values from the simulation approximately
and cannot expect quantitative agreement with the analytical results.
Nevertheless, the analytical results prove useful for understanding
qualitative behavior. Results are shown in Fig. 4 and Table 1.

Similar to cases discussed in paper I, we consider a foil, initially
5 nm thick and having a density of $n_{e0}/n_{c}=1$, and irradiate it by
a drive laser pulse of form $a_{d}\sin (\omega _{i}t)$ with sharp
front and $a_{d}=2$. This is sufficient to drive out all electrons
from the foil and to generate a thin electron layer. It reaches
$\gamma \approx 8$ and $\gamma_x \approx 2$ after two laser cycles
of laser interaction, as it is seen in Fig. 4a. At this point of
time, we let the electron layer interact with a counter-propagating
half-cycle switch pulse (see detailed discussion in paper I) of the
form $-a_{s}\sin (\omega _{i}t)$ with $a_{s}=1.9$. This switch pulse
turns electron momenta into $x$ direction such that $\gamma_x=\gamma
\le 7.9$ for a longer period of time. In this way we have prepared a
relativistic mirror to test the reflectivity $\alpha$.

A 15-cycle flat-top probe pulse is chosen with normal incidence,
$a_{0}=0.01$, and polarized orthogonal to the drive pulse. Both
drive and probe pulse touch the foil at same time $t=0$. The
reflected probe signal is shown in Fig. 4b and its power spectrum in
Fig. 4c. It is detected at a position of $8 \lambda_L$ behind the
foil. In Fig. 4b we see that about 3.5 cycles of probe light are
reflected before the switch pulse hits. In this early phase,
$\gamma_x$ is increasing up to a value of 2.1 . It produces the
chirped part of the reflected pulse, seen for $t/\tau_i \le 8.52$,
and the low-frequency plateau in the spectrum extending up to
$\gamma_x^2(1+\beta_x)^2\approx 16$. Then the switch pulse abruptly
changes the reflected signal. Fast oscillations show up and generate
the high-frequency peak in the spectrum at $\omega/\omega_i \approx
4 \gamma^2 \approx 249$, corresponding to $\gamma\approx 7.9$.

We have extracted the reflectivities from the simulation for the
time intervals before and after switch pulse impact. At this point
of time, the layer density $n_e/n_{e0}=d_0/d$ is estimated from its
thickness $d=6.4$ nm; time-averaged values for $\gamma_x$ and
$\gamma$ are chosen. Results are compared with analytic values in
table 1. One recognizes that Eq. (15) describes the simulated
reflectivities fairly well for two significantly different sets of
mirror parameters. The deviations are within a factor of two and can
be attributed to the averaging over temporal changes in layer
evolution.

\begin{table}
\caption{Fractions of probe light energy reflected before and after
switch pulse. Comparison of simulated results with Eq. (15).}
\label{tab:1}      
\begin{tabular}{lll}
\hline\noalign{\smallskip}
 & before switch & after switch  \\
\noalign{\smallskip}\hline\noalign{\smallskip}
simulation & $5.23\times 10^{-5}$ & $5.55\times 10^{-6}$ \\
analytic (Eq. (15)) & $3.96\times 10^{-5}$ & $3.59\times 10^{-6}$ \\
\noalign{\smallskip}\hline
\end{tabular}
\end{table}

\section{Conclusions}
\label{end}

In conclusion, we have investigated linear coherent Thomson
backscattering of probe radiation obliquely incident on ultra-thin
dense electron layers. They act as relativistic mirrors. In
particular, the present investigation is related to laser
irradiation of ultra-thin foils in a regime in which all electrons
are separated from ions and driven to form the relativistic mirror.
Here an analytical expression for the mirror reflectivity has been
derived and verified by comparison with 1D PIC simulations.

The direction of the coherently reflected light is oriented along
the normal direction of the plane layer, which is also the direction
of the driving laser pulse. Different from incoherent scattering, it
is not oriented along the electron momenta. In the case of direct
laser electron acceleration considered here, electron momentum has
necessarily a transverse component $p_y$ and is inclined relative to
the normal direction. Even tough all these angles become small for
$\gamma \gg 1$, a significant difference remains with respect to
Doppler shifts, which scale $\propto
\gamma_x^2=\gamma^2/(1+(p_y/m_ec)^2)$ rather than $\propto
\gamma^2$. This typically leads to spectra lower in frequency for
coherent scattering, though much higher in intensity. In thin foil
experiments, the transition from coherent to incoherent scattering
may be detected when the layer density decays in time.

The mirror reflectivity derived in this paper is also governed by
$\gamma_x$. For oblique incidence, it is found to be almost independent
of p- or s-polarization; reflectivity rises with angle of incidence,
while frequency falls. The coherent reflectivity oscillates as a function
of layer thickness due to constructive and destructive interference
of light reflected at different layer depth. This modulation is damped
for ultra-short few-cycle pulses due to time delay of reflections from
different depth.

In summary, light reflected coherently from relativistic electron layers
is expected to become an important new source of VUV- and X-radiation.
It will also serve as an excellent tool to diagnose layer electron evolution.

\section*{Acknowledgments}

H.-C. Wu acknowledges support from the Alexander von Humboldt
Foundation. This work was also supported by the DFG project
Transregio TR18, by the Munich Centre for Advanced Photonics (MAP),
and by the Association EURATOM - Max-Planck-Institute for Plasma
Physics.

\end{document}